\documentclass{siamltex1213}

\usepackage{amsmath,amssymb}

\usepackage{color}
\usepackage{graphicx}
\usepackage{subcaption}

\graphicspath{{figs/}}

\newcommand\myurl[2]{\url{#1}} 

\newcommand{\software}[1]{\texttt{#1}}
\newcommand{\libMesh}{\software{libMesh}}
\newcommand{\GRINS}{\software{GRINS}}
\newcommand{\FEMSystem}{\software{FEMSystem}}

\newcommand{\bu}{\boldsymbol{u}}
\newcommand{\bv}{\boldsymbol{v}}
\newcommand{\bp}{\boldsymbol{p}}

\title{GRINS: A Multiphysics Framework Based on the libMesh Finite Element Library}

\author{Paul T. Bauman
  \thanks{Mechanical and Aerospace Engineering,
 Computational and Data-Enabled Science and Engineering,
 University at Buffalo, State University of New York,
 318 Jarvis Hall,
 Buffalo, NY 14260-4400
  (\email{pbauman@buffalo.edu})}
  \and
  Roy H. Stogner
  \thanks{Institute for Computational Engineering and Sciences,
    University of Texas at Austin,
    201 E. 24th St.,
    Austin, TX 78712
  (\email{roystgnr@ices.utexas.edu})}}

\begin{document}
\maketitle
\slugger{sisc}{xxxx}{xx}{x}{x--x}

\begin{abstract}
The progression of scientific computing resources has enabled the
numerical approximation of mathematical models
describing complex physical phenomena. A significant portion of
researcher time is typically dedicated to the development of software
to compute the numerical solutions. This work describes a flexible C++ software
framework, built on the \libMesh{} finite element library, designed to
alleviate developer burden and provide easy access to modern
computational algorithms, including quantity-of-interest-driven parallel adaptive mesh
refinement on unstructured grids and adjoint-based sensitivities. Other
software environments are highlighted and the current work motivated;
in particular, the present work is an attempt to balance software
infrastructure and user flexibility. The applicable class of
problems and design of the software components is discussed in
detail. Several examples demonstrate the effectiveness of the design,
including applications that incorporate uncertainty. Current and planned developments are discussed.

\end{abstract}




\section{Introduction and Motivation}
The evolution of mathematics, algorithms, and software to support the
solution of partial differential equations as well as the deployment
of increasing scientific computing resources have enabled the
study of increasingly complex mathematical models of physical phenomena.
A key aspect of the development of the solution of these
models is the deployment of software elements to effectively compute
solutions using advanced numerical methods and algorithms.
However, the complexity of
both the physical problems being studied and the solution methodologies
used to compute solutions to these problems ensures that the software
development effort will require significant resources in
developer time and software engineering expertise.

For example, a research task may be to calibrate parameters of a
mathematical model using a new numerical formulation, such as a new
stabilization scheme for finite element formulations in fluid
mechanics. As calibration requires many forward solves of the physics
model, efficiency is key.
Thus, one may wish to use adaptive spatial
grids, as well as adaptive and/or high-order time stepping schemes. To drive
adaptive mesh refinement, there are variety of techniques, including
algorithms requiring the solution of an adjoint problem that focus on
reducing errors in functionals of the solution of the mathematical
model, e.g. \cite{BeckerRannacher}.
There are existing software libraries dedicated to supporting
each of these tasks, requiring the developer to become familiar with
each package, bring
the elements together, and deploy a stand-alone program for the study,
e.g. \libMesh~\cite{libMesh}, \software{deal.II}~\cite{BangerthHartmannKanschat2007}, and \software{Trilinos}~\cite{Trilinos}.

While the use of existing libraries is essential, there are still
inefficiencies with focusing on standalone programs for each study.
First, there is little or no reusability with the developed modeling kernels.
When developing the mathematical modeling capability, there may be
opportunities to reuse that model in future work. This is especially
true in multiphysics applications where similar physical phenomena may
occur, e.g. incompressible fluid flow, heat transfer, chemical
tracers, and stabilization operators.
%
In addition to losing the
potential reusability of this physics kernel, one loses the \emph{testing} that
went into that code, e.g. regression and unit testing and MMS-based
testing.
Often substantial developer time goes into testing the
developed kernels and by tying that development to a deployed
application, this effort will need to be duplicated to deploy a
separate stand-alone program that uses the developed physics. Although
one may simply, in principle, ``copy-and-paste'' into the new
application, there will inevitably be modifications such as matching APIs for the
new application, changes to coding style, and adapting to possibly
different testing frameworks. All these changes incur development
costs that could potentially be avoided by using a common
infrastructure in which to develop mathematical models.

Another negative aspect of stand-alone applications is the limited
ability to compare to alternative algorithms and methodologies. If, for example,
one aspect of the developed application is a new finite element
formulation for a particular physics kernel, it would be of interest
to compare and contrast this new formulation with existing
formulations. A framework which easily accommodates the extension of
physics kernels would more easily enable such comparisons as opposed
to stand-alone applications, where either another application or a
framework for handling the multiple kernels would need to be
developed. These ideas have been an important aspect of the
development of ``composable solvers'' within the \software{PETSc} library for linear and
nonlinear solver algorithms~\cite{PETScFAS,PETScLinearSolvers}.

On the other hand, while software frameworks greatly aid in
alleviating the aforementioned concerns, one also potentially loses
one of the advantages of stand-alone software applications:
flexibility for experimentation of algorithms to meet demands for the
problem at hand. This is perhaps the greatest challenge in developing
such frameworks to tackle complex multiphysics applications utilizing
sophisticated numerical strategies, such as adjoint-based adaptive
mesh refinement (AMR),
adaptive modeling,
sensitivity analysis, and, eventually, enabling uncertainty quantification.

%
%
%
The focus of the present work is the development and deployment of a
framework, called \GRINS,\footnote{\GRINS{} stands for ``General Reacting
Incompressible Navier-Stokes''. The original impetus for \GRINS{}
development was reusing incompressible reacting flow kernels. The
development led to a more general framework, but the acronym stuck.}
to support multiphysics finite element applications, the
reusability and extensibility of mathematical modeling kernels, supporting
interfaces to existing solver and discretization
libraries to enable modern solution strategies, while, at the same
time, retaining flexibility to effectively tackle the science or
engineering problem of focus.

The remainder of the paper is devoted to discussing the
underlying libraries used and the description of the \GRINS{} framework.
Alternative libraries and frameworks are compared and
contrasted in the Section~\ref{sec:compare}. Section~\ref{sec:libmesh}
discusses the \libMesh{} finite element library that is the foundation
of the present work. Section~\ref{sec:femsystem} discusses the
\FEMSystem{} framework in \libMesh{} that the current work extends into a
multiphysics framework. Section~\ref{sec:grins} discusses \GRINS{} in
some detail, highlighting the mechanisms that promote reusability and
extensibility in multiphysics applications. Section~\ref{sec:examples}
illustrates several examples that use the \GRINS{} infrastructure before
discussing on-going and future efforts in Section~\ref{sec:conclusions}.

\subsection{Existing Libraries and Frameworks}~\label{sec:compare}
In this section, we briefly highlight software environments with
related goals to the present work. We note that many of
the packages/frameworks discussed below, including the present work,
all started gaining development momentum around 2010; aspects of each
started earlier, but the frameworks seemed to gel around this time. In
a sense, all these packages, including \GRINS, are exploring
trade-offs between ease of development and user flexibility.

\paragraph{MOOSE}

The closest existing software infrastructure to the current work may
be the
Multiphysics Object-Oriented Simulation Environment (\software{MOOSE})~\cite{MOOSE}.
\software{MOOSE} is developed at the Idaho National Lab targeting nuclear scientists
and engineers in support of their modeling and simulation efforts and
currently supports dozens (approaching hundreds) of applications. \software{MOOSE}
is built on \libMesh, as is the current work.  \software{MOOSE} targets scientists and
engineers who may have little software development training but who
wish to rapidly deploy programs aimed at the simulation of nuclear
engineering applications.

One significant area where the current work differs from \software{MOOSE} is
the availability in \GRINS{} of adjoint-based methods for error
estimation, AMR, and sensitivity analysis.
Adjoint-based methods, highlighted in Section~\ref{sec:adjoints}, provide
the opportunity for adaptive mesh refinement based on local functionals
called quantities of interest. Furthermore, they provide a more efficient
means for computing parameter sensitivities in quantities of interest.
\GRINS{} uses \libMesh{} to automatically provide discrete
adjoint computations based on the given Jacobian representation.

\paragraph{FEniCS}

The \software{FEniCS} python package (see~\cite{FEniCSBook} and references therein)
uses a domain specific language (DSL)~\cite{FEniCSUFL} approach to
provide a platform for enabling PDE
solutions based off of the weak forms supplied by the user in the
DSL. This functionality includes the construction of discrete adjoint
operators~\cite{FEniCSUnsteadyAdjoint} and their
solution based on the forward operator supplied.
Assembly, solver interfaces, and other aspects of the simulation are then handled
by the underlying \software{FEniCS} infrastructure. \software{FEniCS} provides a
powerful framework for enabling rapid development of PDE solutions,
but one is also beholden to this framework and user requirements outside
the ecosystem can be challenging to realize. For example, if one is
exploring formulations of the adjoint problem that do not correspond
to the discrete adjoint of the forward operator, i.e. the adjoint of the
forward problem lacks consistency. Such issues arise in many
stabilization schemes in finite element methods of fluid mechanics~\cite{WildeyPaper},
for example. Consider as well the solid mechanics example
in Section~\ref{sec:solids_ex} --- to the knowledge of the authors, this is not
possible to deploy within \software{FEniCS} due to the restriction of the
language distinguishing gradients in physical coordinates
vs. gradients in reference element coordinates as well as supporting
meshes containing elements of different dimension.

\paragraph{Albany}

The \software{Albany} package~\cite{AlbanyWebsite} takes an ``agile components''
approach to their framework, building and connecting dozens of
packages within \software{Trilinos}~\cite{Trilinos} and is beginning to support
user applications~\cite{HansenPaper,AlbanySIAMTalk}. \software{Albany} facilitates interaction with the
different components of \software{Trilinos}: linear solver, nonlinear solver,
optimization, finite element assembly, and other scientific computing
infrastructure. Originally, \software{Albany} interfaced with with the Sierra
toolkit~\cite{Sierra}, but recent efforts seem to indicate collaboration with the
mesh infrastructure of the SCOREC center, see e.g.~\cite{SCOREC-scaling}.
\software{Albany}'s approach is, in a
sense, at a different extreme where instead of a wholistic framework
to which the user much adhere, it endeavors to maximize
reusability of software components. While this can lead to much less
user code to write, it also may hinder determining which lines of code
to write as the user must become familiar with the many different
components to facilitate their simulation requirements.

\paragraph{Other Foundational Packages}

There are several other software packages that provide the foundations
for constructing programs to compute numerical approximations of
complex mathematical models. These include the \software{deal.II}{}
finite element library~\cite{BangerthHartmannKanschat2007},
the \software{DUNE} numerics environment, see e.g.~\cite{DUNE}, and
the \software{OpenFOAM}{} package~\cite{OpenFOAM}. Each of these software
environments provides the framework upon which to construct
application programs targeting a specific problem. In a sense, these
packages, as well as \libMesh{} discussed in Section~\ref{sec:libmesh}, provide the
maximum amount of flexibility for the user, but also provide a higher
barrier-to-entry as there is a more limited amount of infrastructure
upon which to build the application.

%

\section{libMesh Finite Element Library}~\label{sec:libmesh}
The \libMesh{} finite element library~\cite{libMesh} was initiated as part of the
Ph.D. work of Benjamin Kirk~\cite{Kirk2007}, as an alternative to the \software{deal.II}
library~\cite{BangerthHartmannKanschat2007}, in order to provide adaptive mesh
refinement capabilities on
general unstructured meshes, including triangles and tetrahedra.
The library supports a number of
geometric elements including quadrilaterals, hexahedra,
triangles, tetrahedra, and prisms. Supported finite elements include traditional
first and second order Lagrange, both scalar and vector-valued, as
well as arbitrary order hierarchical bases, Clough-Toucher
macroelements, and N\'{e}d\'{e}lec elements of the first type.

\libMesh{} supports mesh partitioning through interfaces to several
packages, including Hilbert space filling curves through libHilbert~\cite{libHilbert} and
graph-based algorithms through Metis and
ParMetis~\cite{METIS}. Additionally, \libMesh{} supports parallel
distributed meshes for
increased scalability and fully supports distributed parallel adaptive mesh refinement.
Currently, \libMesh{} has been scaled tens of thousands of cores and
has been run on over $100{,}000$ cores
on the BG/Q machine Mira at Argonne National Lab~\cite{Gaston_2013}.
A variety of mesh formats are supported to facilitate use of meshes
generated for complex geometries. Additionally, \libMesh{} supports
parallel restart file formats.

More direct \libMesh{} support for multiphysics applications comes via
the existing interfaces to external solver packages,
such as \software{PETSc}~\cite{petsc-web-page,petsc-user-ref,petsc-efficient} and
\software{Trilinos}~\cite{Trilinos}. In particular, the nature of the design of
the \software{PETSc} solvers, as described
in~\cite{Brown:2012:CLS:2411125.2411151}, allows one to experiment
with solver choices, including solver algorithm and preconditioning.
Variables in multiphysics applications can be automatically
assigned to \software{PETSc} field splits by \libMesh{} to enable physics-aware
algorithms.  Next, \libMesh{} supports identifying subdomains within
a mesh. This allows one to easily control which models are
active on particular subdomains to allow, for example, modeling of
conjugate heat transfer and fluid-structure interaction problems. Such
capabilities also enable adaptive modeling algorithms~\cite{BaumanTimo}.
Finally,  \libMesh{} has recently extended the \software{fparser}~\cite{fparser}
library to support both parsing and compilation of mathematical functions
into high performance kernels. This capability allows for easy
specification of boundary conditions, initial conditions, or
constitutive equations from an input file.


\section{FEMSystem Infrastructure}~\label{sec:femsystem}

The \libMesh{} finite element library provides a wide set of tools
with which to build a mesh-based application. However,
\libMesh{} applications were originally required to reimplement many kernels
common to finite element applications, including assembly loops,
time integration schemes.  To reduce the rewriting of these
routines, the \FEMSystem{} infrastructure was
implemented  in the \libMesh{} library as part of the dissertation
work of the second
author~\cite{Stogner:2008vi,Stogner:2008ev,Stogner:2007iq}.
Implementation of the \FEMSystem{} infrastructure began
in 2007, but discussion of the original design is limited to
dissertations of the second author and Dr. John
Peterson~\cite{PetersonThesis}, so we detail the current design here,
both to disseminate to a larger audience and to lay the groundwork for
the design of \GRINS.

\subsection{Problem Class}\label{sec:femsystem_problem_class}

Currently, the \FEMSystem{} infrastructure is designed to accommodate the
following classes of problems. Let $\Omega = \cup \Omega_s$, with
each manifold subdomain $\Omega_s \subset \mathbb{R}^{d_s}$,
$d_s = 0, 1, 2, \text{ or } 3$; problems of mixed dimension are
allowed.  We seek solutions $\bu = (\bu_s)$ for which
each subdomain solution $\bu_s$ is defined at
spatial locations $\mathbf{x} \in \Omega_s$ for times $t \in [0,T)$.
The number and choice of variable components of each Sobolev space to which
$\bu_s$ may belong depends on the subdomain $s$.  For those variables
which span connected subdomains, typically $C^0$ continuity is
enforced, but $C^1$ continuity or discontinuity may be required
instead.  Dirichlet boundary conditions on some $\Gamma_d \subset \cup
\partial \Omega_s$ may also be strongly enforced if specified.

Then, the following system of partial differential equations is
assumed to hold:
\begin{equation}\label{eq:fs_strong_form_1}
\begin{aligned}
M(\bu) \dot{\bu} &= F(\bu), &\text{ in } \Omega \times (0,T) \\
G(\bu) &= 0, & \text{ in } \Omega \times (0,T)  \\
\bu(\mathbf{x},0) &= \bu_0(\mathbf{x})
& \forall \mathbf{x} \in \overline{\Omega} \\
\bu(t) &= \mathbf{g}(t), & \text{ on } \Gamma_d\times (0,T) \\
\mathbf{\sigma}(\bu,t) \cdot \mathbf{n} &= h(t), & \text{ on } \Gamma_n\times (0,T)
\end{aligned}
\end{equation}
where $\dot{(-)}$ indicates a time derivative, $\mathbf{g}(t)$ and
$h(t)$ are
given, $\mathbf{n}$ is the outward unit normal on $\Gamma_n$,
the operator $M$ is the ``mass term'', $F$ is the ``time
derivative term'', and $G$ is the constraint term.  In general each of
these terms and conditions may depend on $\mathbf{x}$ and $t$.
The remainder of the presentation will focus on first order (in
time) systems such as the above, but we note that second order (in
time) systems are also supported within this framework, for which
examples will be shown (Section~\ref{sec:examples}).

The strong form of this system of equations is now cast into a weak form.
In particular, for any test function $\bv \in V$, where $V$ is
an appropriate space of test functions, each term of the strong form
is converted to a consistent variational form:
\begin{equation}\label{eq:first_order_weak_form}
\begin{aligned}
\text{Find } \bu \in L^r(0,T;U):& \\
M\left(\bu, \dot{\bu}; \bv\right)
&= F(\bu; \bv) \quad \forall \bv \in V \\
G(\bu; \bv) &= 0 \quad \forall \bv \in V
\end{aligned}
\end{equation}
where $r$ is an integer that depends on the operators $M, F$, and $G$.
The ``;'' in the semilinear forms indicates that the terms to the left
are possibly nonlinear while the terms to the right are
linear~\footnote{
  While the time derivative is a linear operator, nonlinear
  combinations of time derivative terms do appear in, for example,
  certain stabilized schemes for convection dominated problems. Thus,
  in general, the mass term is possibly nonlinear in the time
  derivative of the state variable.}.
The notation has been abused here: the operator symbols in
the strong form have been reused in the work form; the context will
  distinguish them.
In the present context, the operators
  in~\eqref{eq:first_order_weak_form} are understood to be integral
  operators over $\Omega$ that are typical in the weak form of partial
  differential equations.

Now introduce a tesselation of $\Omega$, $\mathcal{T}^h$, into $N_e$ finite elements
$\Omega_K$ such that
\begin{equation*}
\overline{\Omega} = \bigcup_{K=1}^{N_e} \overline{\Omega}_K, \quad
\Omega_K \cap \Omega_L = \emptyset, \quad \forall K \ne L
\end{equation*}
and define finite element spaces $U^h$ and $V^h$. Then,
problem~\eqref{eq:first_order_weak_form} becomes the following
semi-discrete system:
\begin{equation}\label{eq:first_order_fe_form}
\begin{aligned}
\text{Find } \bu_h \in L^r(0,T;U^h):& \\
M\left(\bu_h, \dot{\bu}_h; v_h\right)
&= F(\bu_h; \bv_h) \quad \forall \bv_h \in V^h \\
G(\bu_h; \bv_h) &= 0 \quad \forall \bv_h \in V^h
\end{aligned}
\end{equation}
Upon introduction of finite element shape functions, i.e.
$\bu_h = \sum_i \bu_i(t) N_i(x)$,
this system of ODEs can now be discretized in
time\footnote{We frame the discussion in terms of the ``method of
lines'', but there's no restriction from the standpoint of the
software framework on following the ``method of Rothe''.}.

Before elaborating on how the software design leverages this
formulation, we introduce an example to help clarify the discussion.
An example of a
system of partial differential equations that can be cast into this
form is the incompressible Navier-Stokes equations:
\begin{alignat}{1}
\rho \dot{\mathbf{u}} = -\mathbf{u} \cdot \nabla \mathbf{u} - \nabla p + \mu \Delta \mathbf{u} + \rho \mathbf{g},
&\text{ in } \Omega \times (0,T) \\
\nabla \cdot \mathbf{u} = 0,& \text{ in } \Omega \times (0,T)
\end{alignat}
where $\mathbf{u}$ is the velocity, $p$ is the pressure, $\rho$ is the
density of the fluid, $\mu$ is the viscosity of the fluid, and $\mathbf{g}$ is
the gravitational vector. These can be recast into weak form,
e.g.~\cite{GunzerbergersBook}. Define
\begin{alignat}{3}
&a: \mathbf{H}^1(\Omega) \times \mathbf{H}^1(\Omega) \rightarrow \mathbb{R}; \quad
&&a(\mathbf{u},\mathbf{v}) = \int_{\Omega} \mu \nabla \mathbf{u} : \nabla \mathbf{v} \; d\mathbf{x} \\
&b: \mathbf{H}^1(\Omega)  \times L^2(\Omega) \rightarrow \mathbb{R};
&&b(\mathbf{v}, q)= -\int_{\Omega} q \nabla \cdot \mathbf{v} \;d\mathbf{x} \\
&c:\mathbf{H}^1(\Omega)\times \mathbf{H}^1(\Omega)\times \mathbf{H}^1(\Omega) \rightarrow \mathbb{R};
\quad
&&c(\mathbf{u},\mathbf{v}, \mathbf{w}) =
\int_{\Omega}
(\mathbf{u}\cdot \nabla)\mathbf{v} \cdot \mathbf{w} \;d\mathbf{x} \\
&f:\mathbf{H}^1(\Omega) \rightarrow \mathbb{R};
\quad
&&f(\mathbf{v}) =
\int_{\Omega} \rho \mathbf{g} \cdot \mathbf{v} \;d\mathbf{x} \\
&m:\mathbf{H}^1(\Omega) \times \mathbf{H}^1(\Omega) \rightarrow \mathbb{R};
\quad
&&m(\mathbf{u}, \mathbf{v}) =
\int_{\Omega} \rho \dot{\mathbf{u}} \cdot \mathbf{v} \;d\mathbf{x} \\
\end{alignat}
For simplicity, we assume $\Gamma = \Gamma_d$ and that the velocity
field is ``no-slip'' on $\Gamma_d$. Define $\mathbf{V}= \mathbf{H}_0^1(\Omega)$
and $Q = L_0^2(\Omega)$.
Then, we seek
$\mathbf{u} \in L^2(0,T;\mathbf{V})$ and
$p \in L^2(0,T;Q)$ which satisfy the following weak form
for almost every $t \in (0,T)$:
\begin{equation}
\begin{aligned}
m(\dot{\mathbf{u}},\mathbf{v}) +
c(\mathbf{u},\mathbf{u}, \mathbf{v}) + a(\mathbf{u}, \mathbf{v}) +
b(\mathbf{v},p) &=
f(\mathbf{v}), \quad &\forall \mathbf{v} \in \mathbf{V} \\
b(\mathbf{u},q) &= 0, &\forall q \in Q
\end{aligned}
\end{equation}

In this case, $\bu = (\mathbf{u},p)$, $\bv = (\mathbf{v},q)$ and clearly
\begin{equation} \label{eq:ns_weak_form}
\begin{aligned}
M\left(\bu,\dot{\bu};
\bv\right) &= m(\mathbf{u},\mathbf{v}) \\
F(\bu; \bv) &= -c(\mathbf{u},\mathbf{u}, \mathbf{v}) - a(\mathbf{u}, \mathbf{v}) -
b(\mathbf{v},p) + f(\mathbf{v}) \\
G(\bu; \bv)  &= b(\mathbf{u},q)
\end{aligned}
\end{equation}
Now, the weak form can be discretized in both space and time,
yielding, in general, a system of nonlinear equations for each time
step. For example, introduce finite element spaces $\mathbf{V}^h$,
$Q^h$, and $U^h=\mathbf{V}^h\times Q^h$. Let
$\mathbf{u}_h \in \mathbf{V}^h$ and $\mathbf{p}_h \in Q^h$.
Assume the resulting system of ODEs are discretized using a variant of the
theta-method. Let the time interval $(0,T)$ be partitioned into a
finite number of segments, $N_t$, and let $n=0,\dots,N_t$. Assume a
uniform timestep length $\Delta t$. Then,
\begin{equation} \label{eq:u_theta}
\begin{aligned}
\bu_h^{\theta} &= \theta \bu_h^{n+1} +(1-\theta)\bu_h^n \\
\dot{\bu}_h &= \frac{\bu_h^{n+1} - \bu_h^{n}}{\Delta t}
\end{aligned}
\end{equation}
Then, given $\bu_h^{n}$ from initial conditions for $n=0$ or from a
previous time step solution, we obtain the following nonlinear system
of equations for $\bu_h^{n+1}$:
\begin{equation} \label{eq:ns_theta}
\begin{aligned}
M\left(\bu_h^{\theta},\dot{\bu}_h; \bv_h\right) &=
F\left(\bu_h^{\theta},\bv_h\right) \\
G\left(\bu_h^{n+1};\bv_h\right) &= 0
\end{aligned}
\end{equation}
Note that the constraint equation is evaluated at the end of the
timestep; this was the impetus for separating the constraint equation
in both the presentation here and in the software framework.

The \FEMSystem{} framework was designed around the following trivial
observations of~\eqref{eq:ns_theta}:
\begin{itemize}
\item As is typical of finite element software, the
residual~\eqref{eq:ns_theta} can be assembled as a sum over element
residual contributions since the integral operator is trivially
decomposed into a sum of integrals over elements.
Thus, the user \emph{only} needs to supply each
of the operators $M\left(\bu_h,\dot{\bu}_h;\bv_h\right)$,
$F(\bu_h; \bv_h)$, and $G\left(\bu_h,\bv_h\right)$ at the element level.
\emph{The user need not know the details of mesh partitioning,
    parallelization, or solver algorithm}.
\item If instead the user wishes to solve only the steady-state
problem, then simply ignoring the $M\left(\bu_h,\dot{\bu}_h;\bv_h\right)$
term in the residual~\eqref{eq:ns_theta} will yield a system of
nonlinear equations for the steady state solution $\bu_h$.
\emph{Thus, the same code can reused for either steady or unsteady
problems.}
\end{itemize}
With a framework that exploits this mathematical structure,
the user only need supply element level calculations for each of the
above operators. All other implementation considerations can be
handled outside the scope of the user and parameters such as
a steady or unsteady formulation can be relegated all the way to the
program input option level. We describe next some of the software design
details that specify such an infrastructure.

We note here that the delegation of residual evaluations is a common
theme amongst the frameworks discussed in
Section~\ref{sec:compare}. The details vary between them: MOOSE uses a
more fine-grained approach,
requiring the user to specify the residual at the degree of
freedom level, i.e. inside quadrature and degree of freedom loops;
FEniCS operates at ``global'' level, where the user is only required
to supply the weak form, e.g.~\eqref{eq:ns_weak_form} in the present context;
Albany uses a template-metaprogramming approach where residuals at each
quadrature point are computed based on residual expressions supplied
by the user. There, the user is required to implement the element, node,
and quadrature loops over the current workload partitioned by the framework.
The present work is an attempt to strike a balance between framework
infrastructure and user flexibility.

\subsection{\FEMSystem{} Design}

This section delves into some details about the software design that
implements the mathematical framework discussed previously. In
particular, we focus on how the element assembly and the solvers are
encapsulated, encouraging code reusability. As with the many other
aspects of \libMesh, the \FEMSystem{} infrastructure makes heavy use
of object-oriented software engineering concepts.

\subsubsection{Element Assembly}

Element assembly of residuals, be they unsteady or steady residuals,
requires the initialization of several components. First, the element
loop must be initiated for the current processor; this is done using
predicated iterators supplied by \libMesh. Additionally, this
abstraction facilitates the use of shared memory parallelism as the
processor local elements can then be divided amongst thread
workers\footnote{Currently, \libMesh{} supports pthreads as well as
Intel Threading Building Blocks~\cite{TBBBook}.}.
Next, for each element, the
element local data must be initialized including element shape
functions evaluated at quadrature points  (and derivatives, and second
derivatives as needed), and solution values for the residual evaluation,
e.g. $u_h^{\theta}$ in~\eqref{eq:u_theta}. All the element local data
is handled by the \software{FEMContext} object. This object is
constructed and passed to the user to facilitate the requirements of
residual evaluation. Using such an object facilitates a consistent API
that can address any model within the problem class described
previously. Again, we emphasize the element level API is independent of the form of the
final residual evaluation. As such, the accumulation of the final residual evaluation can be
delegated to the solver.

\subsubsection{Solvers}

The \software{TimeSolver} hierarchy handles the interaction with the
users supplied residual evaluations. Subclasses
include \software{SteadySolver}, for steady
problems, \software{EulerSolver} that implements the theta method,
and \software{NewmarkSolver} for second order in time systems. Each of
these solvers implement the specific residual form dictated by the
mathematical expressions for each method, delegating the
element-level terms to the user-supplied functions. Note that the
nonlinear solver functionality is delegated to separate objects. This
promotes code reusability and encapsulation of the time stepping away
from the other solver interface considerations. For example,
the \software{EulerSolver} code is only 125 lines.

The nonlinear solver interface is handled by the \software{DiffSolver}
hierarchy. Subclasses include \software{NewtonSolver}, a \libMesh{}
local implementation, and an interface to \software{PETSc}'s SNES
solvers. All \software{DiffSolver} subclasses reuse the \libMesh{} interfaces to
a variety of linear solvers, including \software{PETSc} and \software{Trilinos}.

\subsubsection{User Interaction}

The user can create an application by developing a \software{FEMSystem}
subclass that implements each of the terms, as needed,
in~\eqref{eq:first_order_fe_form}. In particular:
\begin{itemize}
\item $M\left(\bu_h, \dot{\bu}_h; \bv_h\right)$ is implemented
in \software{mass\_residual}
\item $F(\bu_h; \bv_h)$ is implemented
in \software{element\_time\_derivative} for element interior
contributions, \software{side\_time\_derivative} for element boundary
contributions, and \software{nonlocal\_time\_derivative} for scalar
variable contributions, e.g. Lagrange multipliers.
\item $G(\bu_h; \bv_h)$ is implemented
in \software{element\_constraint}, \software{side\_constraint},
and \\ \software{nonlocal\_constraint},
for element interior, element boundary, and scalar variable
contributions respectively.
\item The user can supply Jacobians of the residuals if they are known
analytically or, through the API, the user can indicate that Jacobians
are not computed. In the case Jacobians are not supplied by the user,
they are computed using finite-differences by the \FEMSystem{}
framework. Additionally, an option is provided to use
finite-differenced Jacobians to verify a user-supplied analytic
Jacobian.
\end{itemize}
Additionally, the user must write a \software{main} program that constructs all
the data structures, including the previously discussed \FEMSystem{}
subclass. Several examples are distributed with \libMesh{} and are
available
online\footnote{\url{http://libmesh.github.io/examples.html}}
with the \libMesh{}
documentation\footnote{\url{http://libmesh.github.io/doxygen/index.html}}.

\subsection{Quantities of Interest}\label{sec:adjoints}
In addition to supporting PDEs of the form discussed previously, a
key feature of this work is the support of
computing quantities of interest (QoIs) as well as QoI-based error
estimation and parameter sensitivity computation based on the discrete
adjoint (see e.g~\cite{NocedalWright1999}).
%
%
QoIs are functionals of the solution $\bu$. Therefore, the
user can define the functional
\begin{equation}
Q(\bu): U \rightarrow \mathbb{R}
\end{equation}
Once the QoI is defined, then the process of computing an error
estimate and gradient can be automated using the discrete adjoint.
In particular, for the steady-state case, the linear system for
the discrete adjoint operator
corresponds to the transpose of the Jacobian of the nonlinear system and the forcing
vector is the derivative of the QoI. To elaborate,
rewrite~\eqref{eq:first_order_weak_form} in
a more traditional form. Take $b: U\times V \rightarrow \mathbb R$ and
$f: V \rightarrow \mathbb{R}$ such that $b(\bu;\bv) = F_{n}(\bu;\bv) + G(\bu;\bv)$
and $f(\bv) = F_c(\bv)$ and $F(\bu;\bv) = F_{n}(\bu;\bv) - F_c(\bv)$\footnote{Here, we've
split the operator $F(\bu;\bv)$ into its constant parts (with
respect to $\bu$) and the remaining (nonlinear) part for notational
convenience in order to appeal to the notation of previous
presentations of the adjoint
problem~\cite{PointQoI,OdenPrudhommeModelingError}.}.
Then,
\begin{equation}\label{eq:fwd_problem}
\begin{aligned}
\text{Find } \bu \in U:& \\
b(\bu;\bv) &= f(\bv) \quad \forall \bv \in V
\end{aligned}
\end{equation}
Define the residual $r: U\times V \rightarrow \mathbb{R}$ as
$r(\bu;\bv) = f(\bv) - b(\bu;\bv)$. Define the continuous adjoint solution
$p \in V$ that satisfies the continuous adjoint problem\footnote{There
are alternative definitions of a nonlinear adjoint operator, but the
Fr\'{e}chet derivative-based definition is the most commonly used.}
\begin{equation}\label{eq:adjoint_problem}
\begin{aligned}
\text{Find } \bp \in V:& \\
b^{\prime}(\bu;\bv,\bp) &= Q^{\prime}(\bu;\bv) \quad \forall \bv \in V
\end{aligned}
\end{equation}
where $(-)^{\prime}$ denotes the Fr\'{e}chet derivative. One then
linearizes around the discrete solution $\bu_h$. Assuming an exact
solution of the adjoint problem,
the error representation for the
quantity of interest is~\cite{OdenPrudhommeModelingError}
\begin{equation}
Q(\bu) - Q(\bu_h) = r(\bu_h;\bp) + \Delta
\end{equation}
where $\Delta$ is a term higher order in $\| \bu - \bu_h \|$ and is
typically neglected (and is zero in the case where $r$ is linear in $\bu$).
One must also discretize and solve the adjoint problem, impacting the
error estimate. We defer discussion of the implication of the
discretized adjoint solution to later in this section.

An alternative path to an error representation in the quantity of
interest is to first discretize the forward
problem~\eqref{eq:fwd_problem}:
\begin{equation}\label{eq:discrete_fwd_problem}
\begin{aligned}
\text{Find } \bu_h \in U^h:& \\
b(\bu_h;\bv_h) &= f(\bv_h) \quad \forall \bv_h \in V^h
\end{aligned}
\end{equation}
Now in contrast with~\eqref{eq:adjoint_problem}, the adjoint problem
becomes
\begin{equation}\label{eq:discrete_adjoint_problem}
\begin{aligned}
\text{Find } \bp_h \in V^h:& \\
b_{\bu_h}(\bu_h;\bv_h,\bp_h) &= Q_{\bu_h}(\bu_h; \bv_h) \quad \forall \bv_h \in V^h
\end{aligned}
\end{equation}
where $(-)_{\bu_h}$ denotes the derivative with respect to the discrete
solution $\bu_h$. This ``discretize-then-differentiate'' approach yields
$\bp_h$ and is typically denoted as the discrete adjoint. This approach
is appealing because $b_{\bu_h}(\bu_h;\bv_h,\bp_h)$ is nothing other than the
transpose of the discrete Jacobian of the forward model. Thus, given
the form of the QoI and it's derivative (which can be computed by
finite differences if necessary), then the adjoint solution can be
automated is a physics-independent way. This is exactly the approach
taken in \FEMSystem. Each contribution to the linear
system~\eqref{eq:discrete_adjoint_problem} is then assembled
element-wise in analogy with the assembly of the forward problem as
described in Section~\ref{sec:femsystem_problem_class}.

The theory for QoI-based error estimation has been well
established for some time both in the context of discretization
error~\cite{BeckerRannacher} as well as modeling
error~\cite{OdenPrudhommeModelingError,SISCPaper}.
Given the residual evaluations and adjoint solution, using the
infrastructure described previously,  \libMesh{}
provides several error estimators to drive adaptivity based on this theory.
Currently, there are two QoI-based error estimation
classes: \software{AdjointResidual} which uses (user-specified) global
error estimates in the forward and adjoint solutions from which a
bound on the QoI error can be computed, see e.g.~\cite{VikramDiss} for an
example utilizing \libMesh. Although the effectivity index tends to be
far from unity for such error estimators, they can be effective
error indicators for driving mesh adaptation.

The other core QoI-based error estimation object
is \software{AdjointRefinement} which uses $h$ and/or $p$ refinement
to construct an enriched discretization to compute the adjoint
solution which can then be directly used to produce dual-weighted
residual error estimates.
The issue is that the forward problem has been solved by testing
against all functions in $V^h$, including particularly
$\bp_h$. Several strategies have been proposed, see e.g.~\cite{BeckerRannacher}. Simply
performing a uniform refinement then
transforms~\eqref{eq:discrete_adjoint_problem} to the following
discrete adjoint problem:
\begin{equation}\label{eq:ho2_adjoint_problem}
\begin{aligned}
\text{Find } \bp^{h/2} \in V^{h/2}:& \\
b_{I^{h/2}\bu_h}(I^{h/2}\bu_h;\bv^{h/2}&,\bp^{h/2}) =
Q_{I^{h/2}\bu_h}(I^{h/2}\bu_h; \bv^{h/2}) \quad \forall \bv^{h/2} \in V^{h/2}
\end{aligned}
\end{equation}
where $I^{h/2}$ is an interpolation operator from $\mathcal{T}^h$ to
$\mathcal{T}^{h/2}$. The error estimate is then, ignoring higher order terms,
\begin{equation}
Q(\bu) - Q(\bu_h) \approx r(I^{h/2}\bu_h,\bp^{h/2}) + r(I^{h/2}\bu_h,\bp-\bp^{h/2})
\end{equation}
where various bounds can be used in the latter term; other error
representations are also possible~\cite{OdenPrudhommeModelingError}.
As before, all aspects of the adjoint computation and error estimation
are independent of the actual form of the element contributions and can
be automated by the \FEMSystem{} framework.

Similar ideas hold for the
unsteady case, although the infrastructure to support the unsteady case
in \libMesh~\cite{VikramDiss} is still in preliminary
stages. Currently, the forward solution is stored in memory and no
checkpointing schemes have been employed.

\section{GRINS Framework}~\label{sec:grins}

While \FEMSystem{} provides a platform that eases the development burden
of \libMesh-based applications, the \FEMSystem{} infrastructure still
targets the development of standalone applications. However, as
\libMesh{} is meant to be a library to build finite element
applications, implementations of particular weak forms of mathematical
models do not belong in the library. This is primary impetus
for the \GRINS{} Multiphysics framework: extend the \FEMSystem{} framework
to enable reusable mathematical model kernels and ease the burden in
developing \libMesh-based multiphysics applications. Because the
framework is built atop \libMesh, modern sophisticated parallel
algorithms are intrinsically available.

In addition to adhering to good software development paradigms,
there are two primary
driving goals in the development of \GRINS: \emph{reusability} of the
modules contributed to the framework and ease of \emph{extensibility}
of the modules of the framework. We discuss realizing reusability next
in Section~\ref{sec:reuse} and extensibility in Section~\ref{sec:extend}.


%
\subsection{Reusability}~\label{sec:reuse}
One of the goals of \GRINS{} is to maintain a repository of various
kernels for multiphysics simulation: weak forms (referred to as
\software{Physics} classes in \GRINS), quantities of interest (referred to as
\software{QoI} classes), solver algorithms (\software{Solver} classes), boundary and
initial conditions, and postprocessing. By maintaining this
repository, we can reuse capabilities that have already been developed
and tested and facilitate their reuse to particular applications.
This is accomplished through judicious use of software
design patterns (e.g., see~\cite{DPExplained,GangOfFour}).

The capability to reuse weak form kernels is enabled by
defining an interface to which each weak form will adhere --- this is
done in the \software{Physics} class where the interface borrows heavily
from the \FEMSystem{} interface specification. Each weak form is a subclass of
the \software{Physics} class. Then, in a \FEMSystem{} subclass,
called \software{MultiphysicsSystem}, we implement a strategy pattern
to accumulate contributions to each term in~\eqref{eq:first_order_fe_form} from
each of the enabled \software{Physics} kernels.
In particular, each of the terms can be decomposed into a sum of
$N_p$ operators:
\begin{equation}
\begin{aligned}
\sum_{p=1}^{N_p} M_p\left(\bu_h, \dot{\bu}_h; \bv_h\right) &= \sum_{p=1}^{N_p}F_p(\bu_h; \bv_h) \quad \forall \bv_h \in V^h \\
\sum_{p=1}^{N_p}G_p(\bu_h; \bv_h) &= 0 \quad \forall \bv_h \in V^h
\end{aligned}
\end{equation}
The selection of the
\software{Physics} kernels to be accumulated is parsed from an input file at runtime and
is managed using a factory pattern (\software{PhysicsFactory}).
Similar design strategies were used for the other modules in \GRINS:
\begin{itemize}
\item QoI: Interface defined by \software{QoIBase}, construction done in
  \software{QoIFactory} to enable runtime selection capability
\item Solver: Interface defined by \software{Solver}, construction done in
  \software{SolverFactory} to enable runtime selection capability
\item Postprocessing: Interface defined by
  \software{PostProcessedQuantities}, construction done in
  \software{PostprocessingFactory} to enable additional
  postprocessing of solution variables
\end{itemize}
The interaction between these modules, as well as interactions with
the \libMesh{} library are handled using a \software{Simulation}
object.

By building up reusable kernels of \software{Physics},
\software{Solver}, \software{QoI}, and \software{Postprocessing}
objects, complex multiphysics simulation capabilities can be
constructed with the user supplying only an input file and a mesh (if
the geometry is complex). Thus, if the modeling capabilities required
by the user exist in the framework, then the user can use the input
file and mesh and directly call the \software{grins} executable built
as part of the distribution.

Although a user can use \GRINS{} as a
library and build their own application by extending the various
modules, as discussed in the next section, it is our hope that a
community can be formed that will help build up this repository of
modeling capabilities to facilitate reuse of the different kernels,
thereby eliminating a great deal of code duplication and testing.


%
\subsection{Extensibility}~\label{sec:extend}
While we endeavor to encapsulate a great deal of modeling capability
within \GRINS, we realize that it cannot possibly meet all needs all
the time. Furthermore, the application may have requirements that restrict
dissemination due to security classifications, non-disclosure agreements,
or export control constraints. To this end, \GRINS{} has also been designed with
flexibility and extensibility in its capabilities in order to provide
some ease to the software development burden of enriching the modeling
capability needed by the user while, at the same time, allowing for
reuse of the modeling capabilities present in the framework. This is
accomplished through object-oriented design and use of well
established design patterns. We briefly highlight the extension of
each of the modules below.

\paragraph{Physics}

Extension of the core weak forms available requires creating a
subclass of the \software{Physics} object and overriding the
appropriate methods:\\ \software{element\_time\_derivative}
and \software{element\_constraint} for element interior integrals
and \software{side\_time\_derivative} and \software{side\_constraint}
for boundary contributions that cannot be handled through the boundary
conditions interface (e.g. element jump contributions from
discontinuous Galerkin formulations). The \software{mass\_residual}
function handles contributions to the unsteady terms.
Default ``no-op'' implementations are present in
the \software{Physics} class so that only the methods that are
required by the formulation need be overridden. There are
other functions that can be overridden to promote efficiency, but this is
not required. Once this new \software{Physics} subclass is defined,
then the \software{PhysicsFactory} should be updated so that it can be
invoked from an input file.

\paragraph{Solver}

Although \GRINS{}, through \libMesh, provides basic steady and unsteady solvers,
including mesh adaptive versions, it is conceivable that the user has
a specialized algorithm to use or with which they want to
experiment, such as parameter continuation. To enable such solvers, one
must create a subclass of \software{Solver} and implement its
initialization and the \software{solve} method. The \software{solve}
method is where the user can supply their specialized
algorithm. A \software{SolverContext} is supplied to
the \software{solve} method thereby giving the user access to objects
and options needed. Once the \software{Solver} subclass is
implemented, its creation should be added to
the \software{SolverFactory} to enable runtime selection capabilities.

\paragraph{QoIs}

To create a new QoI, a subclass of \software{QoIBase} must be created
that implements the QoI functionality. The subclass must define the
methods \\ \software{assemble\_on\_interior}
and \software{assemble\_on\_sides} to indicate if the QoI is an
interior, boundary, or both. Additionally, element interior
(\software{element\_qoi} and \\
\software{element\_qoi\_derivative}) and
boundary (\software{side\_qoi} and \software{side\_qoi\_derivative})
assemblies should be overridden as needed. Finally,
the \software{QoIFactory} should be updated to enable the new QoI.





%
\subsection{Other Software Elements}~\label{sec:software}
\GRINS{} is released under an LGPL 2.1 license in order to promote use
by the community. It is developed within the GitHub environment and as
such can be cloned by anyone. It is a C++ framework, but we have not
yet mandated C++11 as a requirement, although we use it when it is
available. Optional thermochemistry libraries are supported,
including Cantera~\cite{Cantera} and Antioch~\cite{Antioch}. Documentation is made
using Doxygen. A unit and regression test suite is included
and integrated with the build system.

\section{Examples}~\label{sec:examples}
We now describe several examples illustrating the reusability of the
existing modeling kernels, QoI-based adaptivity, and interfaces with
the \software{QUESO} statistical library~\cite{QUESO} to facilitate statistical inverse
problems on complex data-reduction modeling problems. We proceed
informally, only discussing the models at a coarse level, to promote
brevity of the manuscript while illustrating the current capabilities
and applications of \GRINS.

\subsection{Fluid Mechanics}

We first illustrate the ability to reuse developed modeling
capabilities with several typical examples from fluid mechanics.
The first two examples share common modeling kernels for which the
input specification\footnote{\GRINS{} uses a fork of the GetPot input
parser~\cite{GetPot} that is distributed with \libMesh.
Particularly desirable features include the
sectioning and the ability to ``include'' other files to facilitate
input file reuse.} is shown in Figure~\ref{fig:ins_common_input}.
\begin{figure}[ht]
\begin{verbatim}
[Physics]
    enabled_physics = 'IncompressibleNavierStokes'
    [./IncompressibleNavierStokes]
       V_FE_family = 'LAGRANGE'
       P_FE_family = 'LAGRANGE'

       V_order = 'SECOND'
       P_order = 'FIRST'

       rho = '1.0' #Density = 1 ==> mu = 1/Re
       mu  = '1.0e-3'
[]
\end{verbatim}
\caption{Commonality in the input file between the lid-driven cavity
         and backward facing step examples.}
\label{fig:ins_common_input}
\end{figure}
\paragraph{Lid-Driven Cavity}

The geometry for the lid-driven cavity is simple enough that the mesh
can be constructed by \libMesh. Thus, all that remains to be specified
are the mesh generation options and the boundary conditions; these are
shown in Figure~\ref{fig:lid_driven_cavity_input}.
\begin{figure}[ht]
\begin{verbatim}
# These options belong in the following sections
#[Physics]
#    [./IncompressibleNavierStokes]
       # Boundary ids:
       # bottom -> 0
       # top    -> 2
       # left   -> 3
       # right  -> 1
       bc_ids = '2 3 1 0'
       bc_types = 'prescribed_vel no_slip no_slip no_slip'

       bound_vel_2 = '1.0 0.0 0.0'
#[]

# Mesh related options
[Mesh]
   [./Generation]
      dimension = '2'
      element_type = 'QUAD9'
      x_min = '0.0'
      x_max = '1.0'
      y_min = '-1.0'
      y_max = '0.0'
      n_elems_x = '15'
      n_elems_y = '15'
[]
\end{verbatim}
\caption{Boundary conditions and mesh generation for the lid-driven
cavity example. For the lid-driven cavity, we supply the velocity at
the top boundary (id = 2), with no slip boundary conditions at all
other boundaries.}
\label{fig:lid_driven_cavity_input}
\end{figure}
There are, of course, many other options that are used to control
other aspects of the calculation such as the
solver (that can be overridden at the command line when using
\software{PETSc}-based solvers), visualization
output, and terminal output. All of these have sensible defaults so
the information given is enough to define a simulation and proceed.

For the given parameters in this example, the convective scales may
not be adequately resolved to yield a stable solution. Thus, one may
wish to add stabilization. There are several stabilization schemes
already present in \GRINS, but should one wish to implement a method
not present, one only need add a new \software{Physics} subclass, as
discussed in Section~\ref{sec:extend}.
\begin{figure}[ht]
\begin{verbatim}
# These options belong in the following sections
#[Physics]
    enabled_physics = 'IncompressibleNavierStokes
                       IncompressibleNavierStokesAdjointStabilization'
#[]
\end{verbatim}
\caption{Enabling stabilization by indicating
additional \software{Physics} class in the \software{enabled\_physics}
argument. Adjoint stabilization means the stabilization operator is
derived from the (minus of the) adjoint of the forward operator.}
\label{fig:ins_stab}
\end{figure}

\paragraph{Backward Facing Step}
The backward facing step example requires a slightly more involved
mesh specification, particularly the association of boundary ids with
specific parts of the boundary, so this mesh was generated externally
and will be read at runtime.
\begin{figure}[ht]
\begin{verbatim}
# These options belong in the following sections
#[Physics]
#    [./IncompressibleNavierStokes]
       # Boundary ids:
       # 1 - Inlet
       # 2 - no slip walls
       # 3 - outlet
       bc_ids = '1 2'
       bc_types = 'parabolic_profile no_slip'

       # u = -480.0*y^2 + 240.0*y = 240.0*y*(1.0 - 2.0*y)
       # v = 0.0
       parabolic_profile_coeffs_1 = '0.0 0.0 -480.0 0.0 240.0 0.0'
       parabolic_profile_var_1 = 'u'
       parabolic_profile_fix_1 = 'v'
#[]

# Mesh related options
[Mesh]
   [./Read]
      filename = 'mesh.e'
[]
\end{verbatim}
\caption{Boundary conditions and mesh generation for the backward
facing step example. In this case, a parabolic profile is specified at
the inlet, no slip on the outer walls, and a natural outlet (hence ``do
nothing'' for boundary 3).}
\label{fig:bfs_input}
\end{figure}
As with the lid-driven cavity, we need to add stabilization to
stabilize the convection scales so we add stabilization as in
Figure~\ref{fig:ins_stab}.

\paragraph{Other Examples}

\begin{figure}[ht]
\begin{center}
\subcaptionbox{Simulation of thermally induced vortex using incompressible
Navier-Stokes, heat transfer, Boussinesq buoyancy, and
stabilization \software{Physics} operators. Completely specified by
user-supplied mesh and input file. Figure provided by Mr. Nicholas
Malaya. \label{fig:sov}}[0.45\textwidth]{\includegraphics[width=0.3\linewidth]{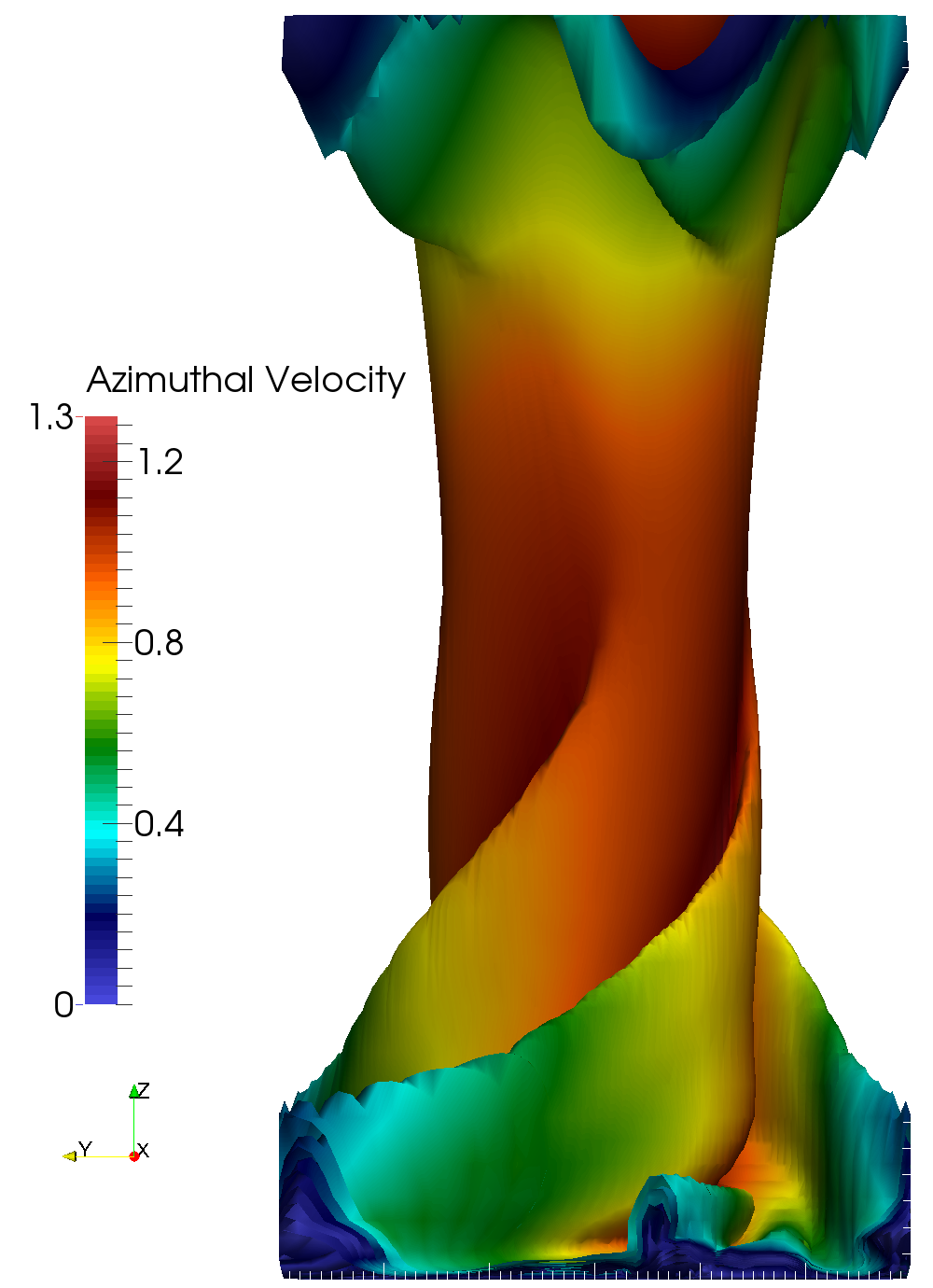}}
\subcaptionbox{Streamlines, colored by temperature, for cavity
benchmark~\cite{CavityBenchmark} problem. Rayleigh number $Ra = 10^8$.
\label{fig:cavity}}[0.45\linewidth]{\includegraphics[width=0.4\linewidth]{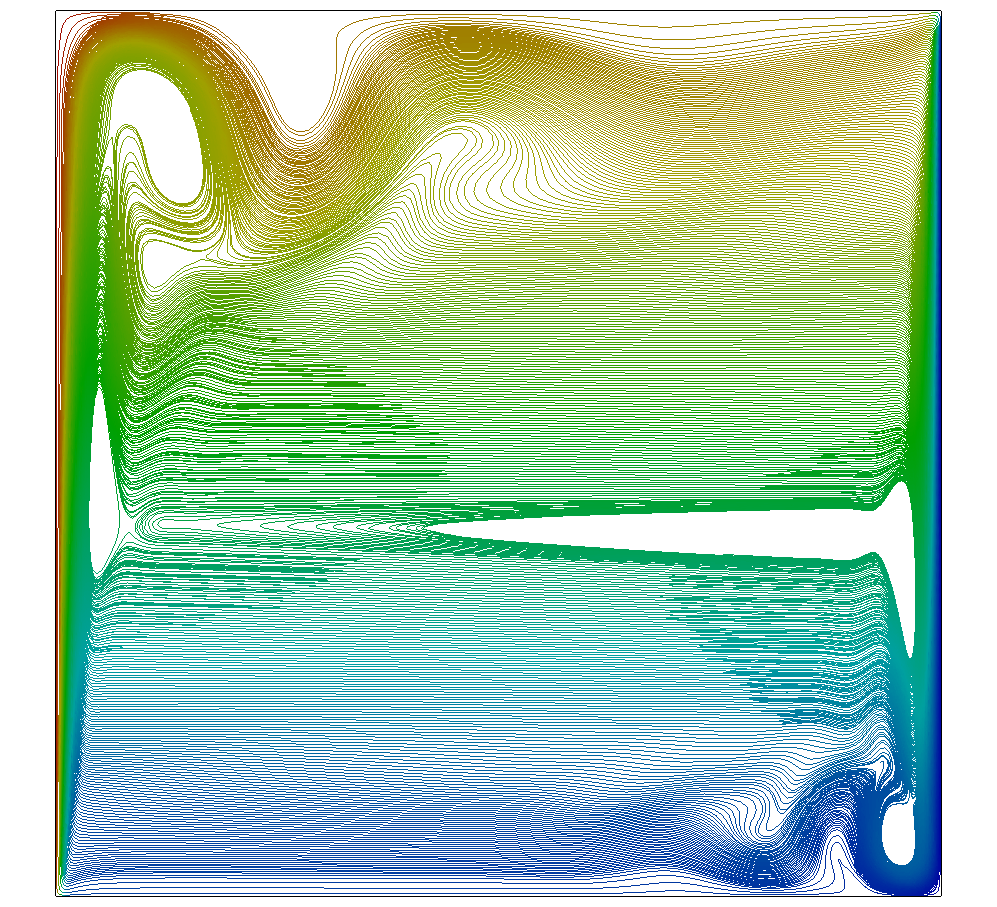}}
\caption{Enabled applications in fluid mechanics.}
\end{center}
\end{figure}

\begin{figure}[ht]
\centerline{\includegraphics[width=0.45\linewidth]{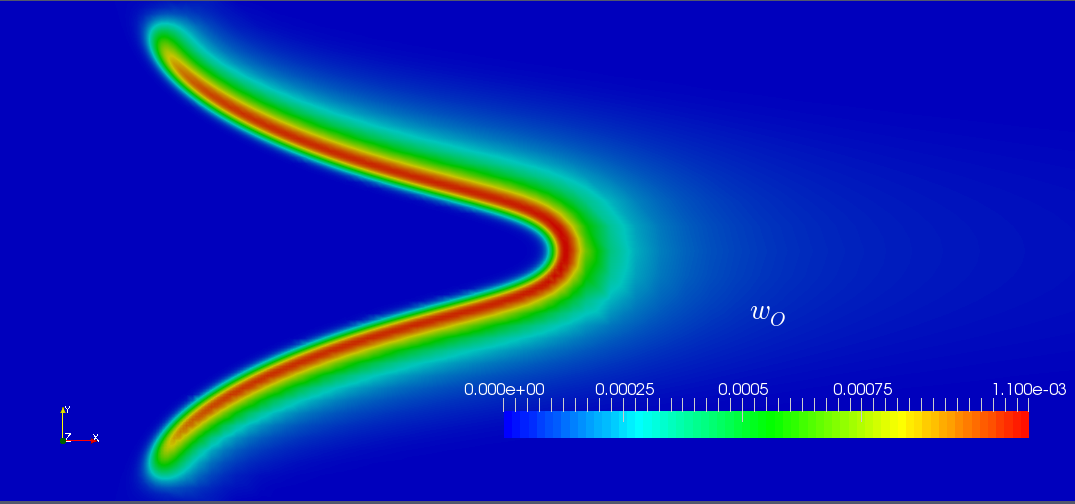}
\includegraphics[width=0.45\linewidth]{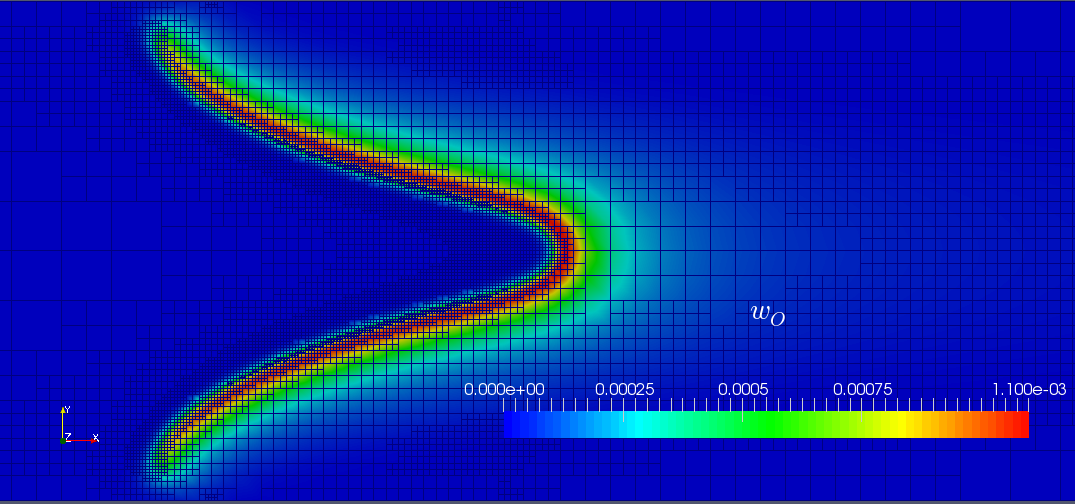}}
\caption{Example of confined Ozone flame, similar to the example given
in Braack et al~\cite{BraackPaper}. Right figure shows mesh adaptation driven by
global error indicators.}
\label{fig:ozone_flame}
\end{figure}

We illustrate more complex examples that benefit from the reusability
of developed \software{Physics} models. First is the study of
thermally induced vortices used for power
generation~\cite{SoV}. Figure~\ref{fig:sov} shows an example completely
specified using a supplied mesh and input file that makes heavy use of
function parsing to construct complex boundary and initial conditions.
Figure~\ref{fig:cavity} shows a solution to a cavity benchmark
problem~\cite{CavityBenchmark} for the Navier-Stokes equations in the
low Mach number limit, see e.g.~\cite{CodinaPaper}. Finally,
Figure~\ref{fig:ozone_flame} shows an example calculation of a
confined Ozone flame, akin to the example in Braack et al~\cite{BraackPaper}.

\subsection{Solid Mechanics}\label{sec:solids_ex}
Now we describe an application in solid mechanics using this
framework. The mathematical models are used to simulate one- and
two-dimensional manifolds of elastic materials that undergo large
deformation, see e.g.~\cite{AlainDissertation}.
Although the final application is for modeling of
parachutes, we show an example that contains many elements of the
desired application whose features illustrate the flexibility of the
current work.

\begin{figure}[ht]
\begin{center}
\subcaptionbox{Perspective view of deformed membrane.\label{fig:sheet_persp}}[0.45\textwidth]{\includegraphics[width=0.4\linewidth]{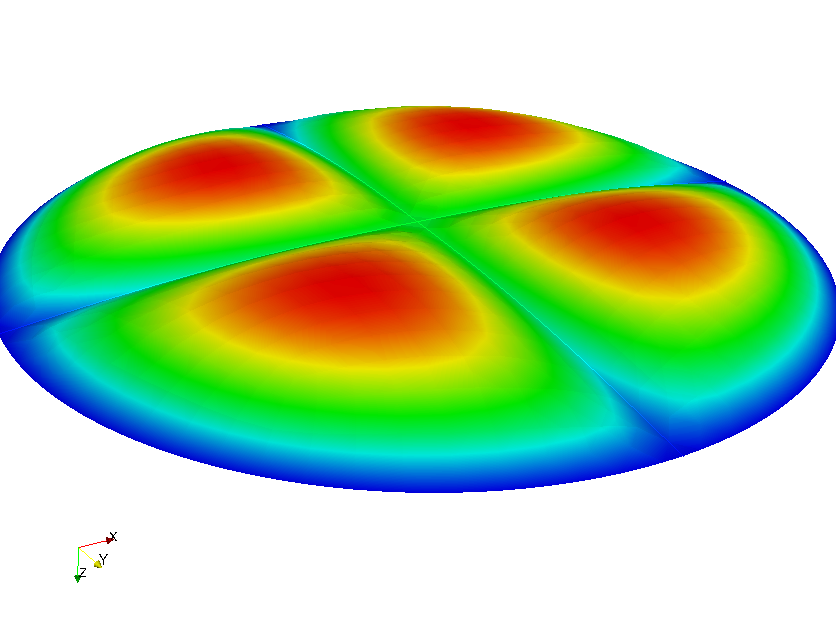}}
\subcaptionbox{Side view of deformed membrane.\label{fig:sheet_side}}[0.45\linewidth]{\includegraphics[width=0.4\linewidth]{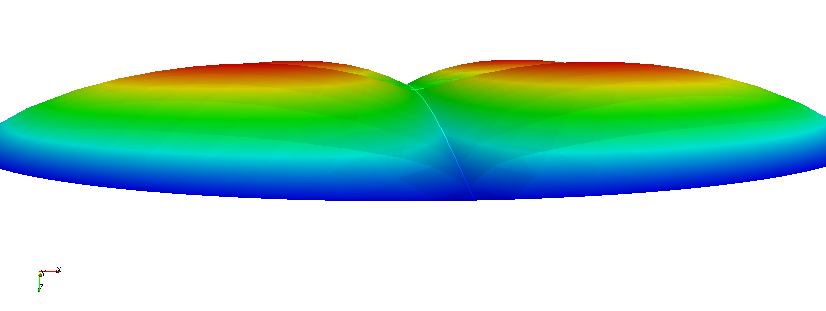}}
\caption{Deformation of two-dimensional rubber membrane with
  one-dimensional axial stiffeners. Rubber material is modeled as a
  Mooney-Rivlin material while the cables are linearly elastic (Hookean).}
\label{fig:inflating_membrane}
\end{center}
\end{figure}

The body is a two-dimensional manifold with  overlapping one-dimensional
``stiffeners'' along with the axial directions and pressure
distribution acting normal to the surface; the latter induces another
geometric nonlinearity in addition to the large deformation. In particular,
Figure~\ref{fig:inflating_membrane} shows a two-dimensional circular rubber membrane
deforming under a uniform pressure loading, acting normal to the
surface, with one-dimensional ``stiffeners'' (cables) along the x- and
y-axis. The rubber membrane is modeled as a Mooney-Rivlin material (material
nonlinearity) and the cables are modeled as linear elastic
materials\footnote{The material is still large deformation, but it is
  assumed that the \emph{strains} are small.}.
This application exhibits multiple materials, multiple dimensional
mesh elements directly coupled, and the underlying nonlinear plane
stress elasticity equations are posed on the manifolds. The latter two
elements, to best of the authors knowledge, are challenging to deploy
within the FEniCS platform\footnote{Because the equations are plane stress
and the reference configuration of the body is not planar in the x-y
plane, then the weak form is posed in convected curvilinear
coordinates such that gradients are with respect to reference element
coordinates and not the more typical physical element coordinates.}.

This example can use both an unsteady solver with heavy
damping or simple continuation solver extension where the applied
pressure is incrementally increased. The latter is distributed as part
of this example in \GRINS{} and required roughly 100 lines of code.

%
%
\subsection{QoI-Driven Mesh Adaptivivty}
This example illustrates using the \software{QoI} and
\software{Solver} infrastructure to enable QoI-based AMR using error
estimates based on adjoint solutions. This particular example
mimics that of Prudhomme and Oden~\cite{PointQoI,PrudhommeDiss} for point-valued QoIs,
namely the Poisson equation with a prescribed loading such that the
solution is a given function.
\begin{figure}[Ht]
\begin{center}
\subcaptionbox{Forward solution.}[0.45\textwidth]{\includegraphics[scale=0.25]{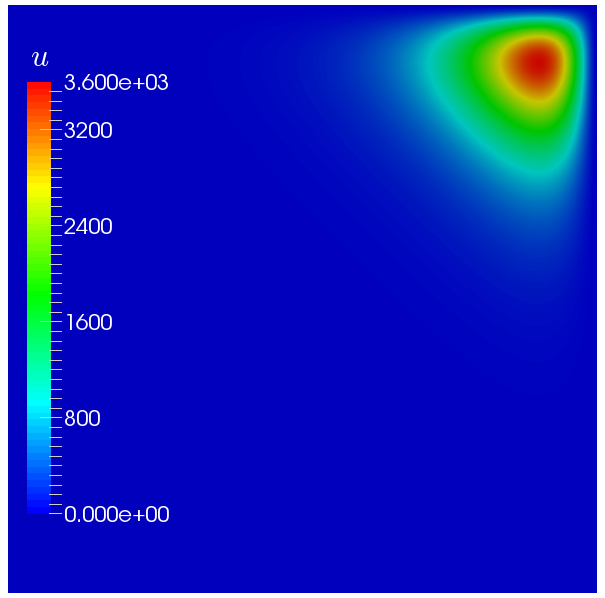}}
\hfill
\subcaptionbox{Adjoint solution.}[0.45\linewidth]{\includegraphics[scale=0.25]{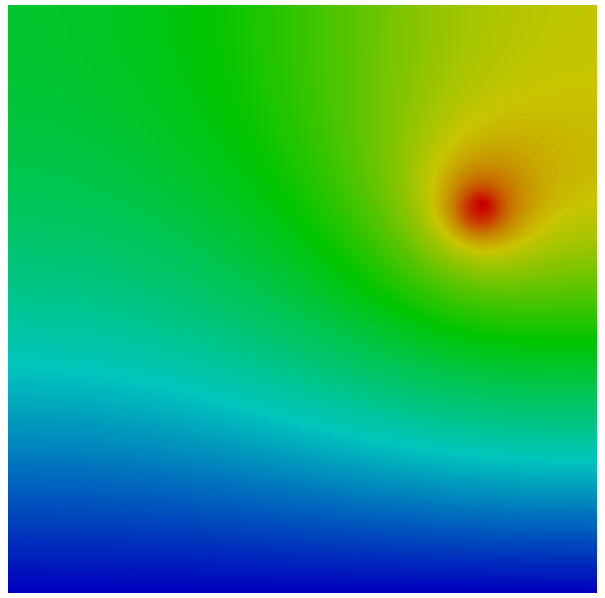}}
\caption{Forward and adjoint solution for $u(0.8,0.65)$ QoI of Poisson
equation.}
\end{center}
\end{figure}
We note that the infrastructure automatically supports multiple QoIs
for both error estimation/adaptive mesh refinement and for parameter sensitivity.

\subsection{Data Reduction Modeling}

The final set of examples that we cite here are the initial
integration with the \software{QUESO} statistical library~\cite{QUESO}
for solving statistical inverse problems. In particular, all of these
examples use \GRINS{} as a \emph{data reduction model} ---
mathematical models are used to infer desired quantities from
experiments given physical measured data and its associated
uncertainty.

The first application was using one-dimensional models of thermocouple gages
to infer surface heat flux~\cite{Thermocouple}. In this case, the thermal
conductivity was unknown and temperature history data was used to
infer the conductivity using a Bayesian approach. Several likelihood models were considered,
all using the underlying forward model infrastructure.
The second application was the real-time inference of a damage field
given given strain field
measurements~\cite{DDDAS_IJNME,DDDAS_CompositesB}. Here, an extended
Kalman filter approach was used to update the damage predictions based
on measured strain data; the underlying forward model in \GRINS{}
supplied to forward model predictions derivatives.
The final application is the inference of surface chemistry
parameters~\cite{Nitridation}. This case uses a two-dimensional reacting flow
model to study the consumption of carbon by dissociated nitrogen from
which surface chemistry between carbon and nitridation can be
inferred. Such reactions are important in reentry flows~\cite{HypersonicsUQpaper}.

At this point, all the previous examples are small application codes
built using \GRINS{} and \software{QUESO}. Current work is
providing direct interfaces to \software{QUESO} that will facilitate
runtime selection of parameters and construction of data reduction
models without the need for separate applications. Completion of this
work will be the subject of future publications.

\section{Concluding Remarks and Future Directions}~\label{sec:conclusions}
In this paper, we have described the \GRINS{} multiphysics framework
that provides a platform for constructing numerical approximations of
complex mathematical models and facilitating the use of modern
numerical methods, including adjoint-based error estimation, AMR, and
interactions with statistical libraries for performing inference.
Users can directly use existing functionality to run calculations only
from an input file and supplied mesh and experiment with the solvers
and parameters to find the right method for the problem; we were
particularly inspired by the \software{PETSc} library for this approach. The
approach taken here is attempt to balance between infrastructure
mandates, e.g. as in \software{FEniCS}, and the flexibility afforded by building
a stand-alone application.

Ongoing work includes building interfaces between \software{PETSc}
and \libMesh{} to facilitate the use of adapted grids for geometric
multigrid based solvers, building reusable interfaces
to \software{QUESO} to facilitate rapid deployment of inference
applications, more flexibility to interfaces to allow user-specified
adjoint operators, infrastructure to support arbitrary
Lagrangian-Eulerian (ALE) fluid-structure
interaction, as well as many other fine-grained enhancements to
lower the barrier for users.

\section*{Acknowledgements}
The authors thank Mr. Nicholas Malaya at the University of Texas at
Austin for providing the thermal vortex image. The first author during
his time in the PECOS center and the second author were supported
the Department of Energy [National Nuclear Security Administration]
under Award Number [DE-FC52- 08NA28615].
The solid mechanics example was developed by the first author under
NASA grant NNX14AI27A. This support is gratefully acknowledged.


\bibliographystyle{siam}
\bibliography{refs}

\end{document}